\documentclass[]{svjour3}  %draft,narrowdisplay,final,reviewcopy,referee
\usepackage{epsfig} % Include figure files
\usepackage{graphicx}% Include figure files
\usepackage{dcolumn}% Align table columns on decimal point
\usepackage{bm}% bold math
\usepackage[usenames]{color}
\usepackage{ulem,bigstrut} %strikeout \sout { }
\usepackage{amsmath}
\usepackage{amssymb}
\usepackage[]{natbib}
\usepackage{subfigure}
\usepackage{ulem}

\def\b{\textcolor{blue}}

\journalname{Bulletin of Mathematical Biology}

\begin{document}

\title{Inferring unobserved multistrain epidemic sub-populations using synchronization dynamics}

\author{Eric Forgoston \and Leah B. Shaw  \and Ira B. Schwartz}

\institute{Eric Forgoston \at
 Department of Mathematical Sciences, Montclair State University, 1 Normal Avenue, Montclair, NJ 07043, USA\\
 \email{eric.forgoston@montclair.edu}
\and Leah B. Shaw \at Department of Applied Science, The College of William \&
Mary, P.O. Box 8795, Williamsburg, VA 23187-8795, USA\\
\email{lbshaw@wm.edu}
 \and Ira B. Schwartz \at  Nonlinear Systems Dynamics Section, Plasma Physics Division, Code 6792,  U.S. Naval Research Laboratory, Washington, DC 20375, USA\\
 \email{ira.schwartz@nrl.navy.mil}
}

\date{Received: date / Accepted: date}

\maketitle

\begin{abstract}
A new method is proposed to infer unobserved epidemic sub-populations by
exploiting the synchronization properties of multistrain epidemic models.  A
model for dengue fever is driven by simulated data from secondary infective
populations.  Primary infective populations in the driven system synchronize
to the correct values from the driver system.  Most hospital cases of dengue
are secondary infections, so this method provides a way to deduce unobserved
primary infection levels.  We derive center manifold equations that relate the
driven system to the driver system and thus motivate the use of
synchronization to predict unobserved primary infectives.
Synchronization stability between primary and secondary infections is demonstrated through numerical measurements of conditional
Lyapunov exponents and through time series simulations.
%\PACS{PACS code1 \and PACS code2 \and more}
\end{abstract}

\section{Introduction}\label{sec:intro}

%\r{General problem of using data to predict epidemics}

Understanding  spread of disease requires both observational
data and mathematical modeling of some class or form \citep{Anderson91}. However, much of
the measured data is quite limited in that it is observed only once during
typically short time intervals and typically is non-stationary. In many instances of observed disease spread,  only case numbers per unit time
are measured. As a result, even for simple, well known diseases, it is difficult
to reconcile the models with the data.

 An important component of epidemic modeling that
cannot be reconciled with data are those individuals required to complete the
 disease path from susceptible to observed infected, which may consist of asymptomatic, individuals. Typically,  intermittent stages along
 the disease path in
 between susceptible and infected cannot be measured directly. For example, in childhood diseases,
certain susceptible individuals may have come in contact with someone who
is infectious, but the resulting infected individual remains latent for a period of
time.  Sub-populations  of such intermittent states  are typically not
observed directly, and must be inferred \citep{Forgoston13}.

Alternative geometric modeling approaches to understanding epidemic spread have
  been examined using  time series analysis from spatio-temporal case observations.  Tools
  from nonlinear time series analysis using embedding theory have been applied
  to measles data~\citep{schaffer93,Blarer1999} to examine chaotic-like
  predictability of case history in the short term.

Another time series method of  analysis  for epidemic spread is that of Time-series
Susceptible-Infected-Recovered (TSIR) modeling~\citep{BFG2002}. Local fits of time-series data generate measures of local
reproductive rates of infection. For childhood diseases, the main assumption is  that in the
pre-vaccine years, all newborns introduced into the population as susceptible
individuals become infected. However, the model excludes any latency period of infection, since it
only considers models of SIR type. Therefore, the model does not predict time
variations in the latent sub-population.

In general, since epidemic data is limited, detailed modeling of disease spread is
required and thus is widespread. Connections of data with full models which are
higher dimensional are difficult since there may exist many unobserved
sub-populations along the disease path. High dimensional coupled patch models of
  cities have limited data time series when compared to the large simulation model dimension. Such limited data sets imply the need for accurate lower
dimensional models to reduce the number of unknown parameters. Latency of infection, which
introduces a series of  exposed classes approximating mean delay times, is
one example that generates high dimensional models. However, it is known
rigorously that the  dynamics in higher dimensional deterministic models
often relaxes asymptotically onto lower dimensional
hyper-surfaces, or center manifolds~\citep{ss83,shbisc07}.  The advantage
in  doing center manifold reductions is that if one can only observe certain components
of a disease, then it is possible to explicitly construct a function that
relates the unobserved components (such as latency, or asymptomatic infections) to
those explicitly measured or observed. Such a relationship, in which
unobserved sub-populations are related dynamically to populations of disease that are measured, is closely aligned to synchronization of coupled systems.

Synchronization has frequently been studied in driver/driven dynamical systems
(also called transmitter/receiver systems), in which
information is transmitted unidirectionally from the driver to the driven
system \citep{BoccalettiKOVZ02}.  Some studies have
focused on controlling parameters in the driven system until synchronization
with the driver is achieved, as a way of determining unknown parameters in the
driver system \citep{Parlitz96,Dedieu97,Chen02,Huang13}.  When the systems are
synchronized, observation of the driven system also allows unobserved
variables from the driver system to be observed 
\citep{Dedieu97}.  In this paper, we propose a similar approach to determine
unobserved quantities during an epidemic.  The driver will be an observed time
series, such as infection prevalence.
%unidirectional coupling of an observed epidemic time series (the driver) to

Our proposed method to infer unobserved epidemic compartments differs from
previous approaches because we exploit synchronization dynamics rather than
using statistical inference techniques.  Previous studies focus more on
integrating over or otherwise accounting for uncertainty in unknown
compartments \citep{Gibson04,Lekone06}.  However, it is frequently seen
in epidemic models that the dynamics approach a lower dimensional center
manifold.  Examples include susceptible-exposed-infected-recovered (SEIR) models
\citep{fobisc09,Forgoston13} and multistrain models for dengue fever
\citep{shbisc07}.  Such results suggest that full information about an
epidemic system is not needed,  and variables that are more important to the
dynamics can be used to deduce other, unknown variables.

We illustrate with a simple example for which it can be shown analytically that driving a system of unknown compartments with values from known compartments will result in convergence to the correct values for the unknown compartments.
Consider a susceptible-infected-recovered (SIR) epidemic in a population with
susceptible, infected, and recovered fractions $s(t)$, $i(t)$, and $r(t)$,
respectively; birth and death rate $\mu$; contact rate $\beta$; and recovery rate $\sigma$.  Suppose that the population dynamics is described by the system
\begin{subequations}
\begin{flalign}
&\dot{s} = \mu - \beta s i -\mu s\\
&\dot{i} = \beta s i-\sigma i-\mu i\\
&\dot{r} = \sigma i - \mu r
\end{flalign}
\end{subequations}
If a time series of the infected fraction $i(t)$ is known (e.g., from public health reporting) but the remaining compartments are unobserved, the susceptible fraction can be determined by driving a new differential equation with the known $i(t)$ as follows.  Let $s_d(t)$ be the driven variable that we hope will reproduce the correct susceptible fraction, and evolve it according to
\begin{equation}
\dot{s}_d = \mu-\beta s_d i -\mu s_d.
\end{equation}
The difference between $s$ and $s_d$, $\xi=s-s_d$, obeys
\begin{equation}
\dot{\xi}=-(\beta i+\mu)\xi.
\end{equation}
Because $\beta i+\mu>0$ for all $t>0$, $\xi$ will approach 0 and $s_d$ will approach $s$ in the long time limit.  Thus, given a sufficiently long time series for $i(t)$, the unknown susceptible levels can be determined for the latter part of the time series.

The application we will study here is a multistrain model for dengue fever
\citep{SSCBMB05}.  Dengue is a mosquito-borne disease with four co-circulating
serotypes.  In previous observations, secondary infections led to more severe
illness and constituted the majority of dengue hospital cases (e.g.,
\cite{nisa03}), so it is plausible that one may have data about secondary
infections and want to deduce other asymptomatic quantities such as primary infective levels.  It has been shown \citep{shbisc07} that the dynamics of this dengue model lie on a lower dimensional center manifold.  In particular, peaks in primary and secondary infective levels are observed to coincide \citep{SSCBMB05}.
%, and it has been proposed that this has been used to infer information about secondary infectives.
However, the equations for secondary infectives derived in \cite{shbisc07}
require knowledge of susceptible and recovered compartments in addition to
primary infectives.  In this paper, we will show that primary infective levels
can be determined by driving a system of equations with known secondary
infective time series. Such a system has the practical value of using real
  measured data to extract unobserved compartments.

In Section \ref{model}, we review the multistrain model for dengue that will
be used in this study.  Section \ref{cm} presents center manifold analysis
showing the relationship between driven and driver variables.  Numerical
results are presented in Section \ref{results} demonstrating the efficacy of
the proposed method, and Section \ref{sec:conc} concludes the article.

\section{The Two-Serotype Model}\label{model}
We begin by describing the compartmental multistrain disease model found in~\citet{SSCBMB05}, studying only
two co-circulating serotypes for simplicity, but we remark that we expect the
  results to hold for arbitrary numbers of strains.  We assume that a given population may be
divided into the following classes that evolve in time:
\begin{enumerate}
\item Susceptible class $s(t)$ consists of the fraction of the total
  population that is susceptible to all serotypes.
\item Primary infectious classes $x_i(t)$ consist of the fraction of the total
  population that is infected with serotype $i$ and is capable of transmitting serotype $i$ to susceptible
  individuals.
\item Primary recovered classes $r_i(t)$ consist of the fraction of the total
  population that has recovered from being infected with serotype $i$.
\item Secondary infectious classes $x_{ij}(t)$ consist of the fraction of the
  total population that is currently infected with serotype $j$, but
  previously was infected with serotype $i$, where $i\neq j$.
\end{enumerate}

In this model, susceptible individuals may develop a primary
infection from either of the two serotypes. Upon recovering, the individual is
immune to the strain that caused the primary infection.  However, the
individual may develop a secondary infection from the second serotype. The infectiousness
of this secondary infection can be increased through an antibody-dependent
enhancement factor.  Upon recovering from the secondary infection, the
individual is immune to both serotypes. A flow
diagram for this model is given in Fig.~\ref{fig:flow}.

\setlength{\unitlength}{2cm}
\begin{figure}%[!ht]
  \begin{picture}(30.,3.)
    \put(0.1,1.1){$\mu$}
    \put(0.3,1.12){\vector(1,0){0.7}}

    \put(1.05,1.1){$\boxed{s}$}
    \put(1.3,1.12){\vector(1,1){.5}}
    \put(1.3,1.12){\vector(1,-1){.5}}
    \put(1.6,1.1){$\beta$}

    \put(1.9,1.7){$\boxed{x_1}$}
    \put(1.9,.55){$\boxed{x_2}$}
    \put(2.25,1.7){\vector(1,0){.7}}
    \put(2.25,.55){\vector(1,0){0.7}}
    \put(2.55,1.9){$\sigma$}
    \put(2.55,.25){$\sigma$}
%    \qbezier(4.7,2)(5.9,2.6)(7.25,1.95)
%    \qbezier(4.7,.3)(5.9,-.3)(7.25,.36)
%    \put(7.3,1.9){\vector(1,-1){0}}
%    \put(7.3,.4){\vector(1,1){0}}

%    \put(5.5,2.6){$\beta(1-\epsilon$)}
%    \put(5.5,-0.4){$\beta(1-\epsilon$)}

    \put(3.1,1.7){$\boxed{r_1}$}
    \put(3.1,.55){$\boxed{r_2}$}
    \put(3.5,1.7){\vector(1,0){0.7}}
    \put(3.5,.55){\vector(1,0){0.7}}
    \put(3.8,1.9){$\beta$}
    \put(3.8,.25){$\beta$}

    \put(4.4,1.7){$\boxed{x_{12}}$}
    \put(4.4,.55){$\boxed{x_{21}}$}
    \put(4.9,1.7){\vector(1,-1){0.5}}
    \put(4.9,.55){\vector(1,1){0.5}}
%    \put(6.4,1.9){$\beta$}
%    \put(6.4,.25){$\beta$}

%    \put(7.1,1.7){$\boxed{x_{12}}$}
%    \put(7.1,.55){$\boxed{x_{21}}$}
%    \put(7.6,1.7){\vector(1,-1){0.5}}
%    \put(7.6,.55){\vector(1,1){0.5}}
    \put(5.0,1.1){$\sigma$}
    \put(4.7,2.0){$\phi$}
    \put(4.7,.15){$\phi$}

    \put(5.55,1.1){$\boxed{r_{tot}}$}
  \end{picture}
  \vspace{.5cm}
\caption{Flow diagram for a multistrain disease model with two co-circulating
  serotypes~\citep{SSCBMB05}. Included is the possibility of enhancing the
  secondary infectiousness through an antibody-dependent enhancement factor $\phi$.} \label{fig:flow}
\end{figure}
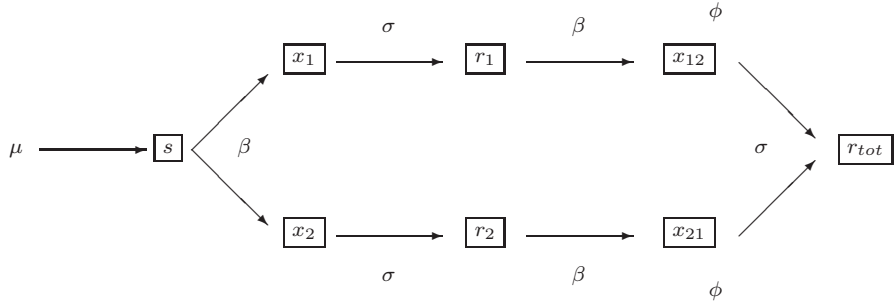

The governing equations for the two serotype multistrain disease model are
\begin{subequations}
\begin{flalign}
&\dot{s} = \mu - \beta s (x_1+x_2+\phi(x_{21}+x_{12})),\label{e:s}\\
&\dot{x}_1 = \beta s (x_1+\phi x_{21})-\sigma x_1,\\
&\dot{x}_2 = \beta s (x_2+\phi x_{12})-\sigma x_2,\\
&\dot{r}_1 = \sigma x_1 - \beta r_1(x_2+\phi x_{12}),\\
&\dot{r}_2 = \sigma x_2 - \beta r_2(x_1+\phi x_{21}),\\
&\dot{x}_{21} = \beta r_2(x_1+\phi x_{21})-\sigma x_{21},\\
&\dot{x}_{12} = \beta r_1(x_2+\phi x_{12})-\sigma x_{12},\label{e:x12}
\end{flalign}
\end{subequations}
where $\mu$ represents a constant birth rate, $\beta$ is the contact rate,
$\phi$ is a constant that determines the antibody-dependent enhancement (ADE),
and $\sigma$ is the rate of recovery, so that $1/\sigma$ is the mean
infectious period.  Although the contact rate $\beta$ could be given by a
time-dependent function (e.g. due to seasonal fluctuations in the mosquito
vector population), for simplicity, we assume $\beta$ to be constant. Unlike
single strain models of $SIR$ type with constant contact rate, 
Eqs.~(\ref{e:s})-(\ref{e:x12}) possess a range of $\phi$ where the endemic equilibrium is unstable.

Rates of infection due to primary infectious individuals have the form $\beta
sx_i$, as found in a classical SIR epidemiological model.  However,
rates of infection due to secondary infectious individuals are weighted by the
ADE parameter $\phi$, and have the form $\beta\phi sx_{ij}$.  If $\phi = 1$,
then there is no antibody-dependent enhancement, and the primary and secondary
infectious individuals are equally infectious. If $\phi = 2$, then secondary
infectious individuals are twice as infectious as primary infectious
individuals, and so forth.  As long as $\phi > 1$ the nonlinear
terms involving secondary infectious individuals will contain an
antibody-dependent enhancement factor.

Throughout this article, we use the following parameter values: $\mu=0.02
$(years)$^{-1}$, $\beta=200$(years)$^{-1}$, $\phi=3$(years)$^{-1}$, and
$\sigma=50$(years)$^{-1}$.  These disease parameters are consistent with
estimates previously used in modeling dengue fever and are summarized in
Table~\ref{tab:parameters}.  In particular, the contact rate $\beta$
corresponds to a reproductive rate of infection $R_0$ of 3.2-4.8, which is
consistent with estimates found in~\citet{fedoan99} and~\citet{nagkoe08}.

The antibody-dependent enhancement parameter $\phi$ is selected to put the
system in the chaotic regime in which the strains are desynchronized, which is
thought to be the biologically relevant dynamics
(c.f. \cite{cummings,shbisc07}).  Dengue has four serotypes, but we model only
two here for simplicity.  The dynamics are similar for two and four serotypes,
although there are shifts in the locations of bifurcation points and thus of
realistic $\phi$ values \citep{Billings07}.  We anticipate that all the
qualitative results of this paper will hold for more realistic four-serotype
models.

\begin{table}
\caption{Model parameters.}

%\begin{center} \footnotesize
\begin{tabular}{p{5.6cm} p{2.12cm} p{3.2cm}} \hline
 \noalign{\vskip 2mm}
 Parameter & Value & Reference \\
\noalign{\vskip 2mm}
\hline
\noalign{\vskip 2mm}
$\mu$, birth rate, years$^{-1}$ & $\sim 0.02$ &  \cite{fedoan99}\\
\noalign{\vskip 2mm}
$\beta$, transmission coefficient, years$^{-1}$ & $\sim 200$ & \cite{fedoan99}\\
\noalign{\vskip 2mm}
$\phi$, ADE parameter, years$^{-1}$ & $\geq 1$ & \cite{SSCBMB05}\\
\noalign{\vskip 2mm}
$\sigma$, recovery rate, years$^{-1}$ & $50$ & \cite{Rigau98}\\
\noalign{\vskip 2mm}
$n$, number of strains & $2$ & -\\
\noalign{\vskip 2mm}
\hline
\end{tabular}
%\end{center}
\label{tab:parameters}
\end{table}

It should be noted that mortality terms have been omitted from
Eqs.~(\ref{e:s})-(\ref{e:x12}).  In the analysis that follows, it is useful to analytically determine the endemic steady state.  This equilibrium
state is not easy to find analytically when mortality terms are
included, but this state is close to the one found when all mortality occurs after recovery from infection with two serotypes, and the mortality rate of the other compartments is $\mu_m = 0$.  Previous work~\citep{shbisc07} has shown that the dynamics with
$\mu_m =0$ are qualitatively similar to the dynamics (and have the same
bifurcation structure) when the mortality rate
$\mu_m$ is equal to the value of the birth rate $\mu=0.02$ used in this
article.  Furthermore, the $\mu_m =0$ assumption is physically reasonable
since the mortality rate for dengue is low and the average age at infection is
believed to be young~\citep{nisa03,shbisc07}

The governing equations for the two serotype multistrain disease subsystem
that are driven by the secondary infectious individuals of
Eqs.~(\ref{e:s})-(\ref{e:x12}) are
\begin{subequations}
\begin{flalign}
&\dot{s}_d = \mu - \beta s_d (x_{d1}+x_{d2}+\phi(x_{21}+x_{12})),\label{e:sd}\\
&\dot{x}_{1d} = \beta s_d (x_{1d}+\phi x_{21})-\sigma x_{1d},\\
&\dot{x}_{2d} = \beta s_d (x_{2d}+\phi x_{12})-\sigma x_{2d},\label{e:xd2}\\
&\dot{r}_{1d} = \sigma x_{1d} - \beta r_{1d}(x_{2d}+\phi x_{12}),\\
&\dot{r}_{2d} =  \sigma x_{2d} - \beta r_{2d}(x_{1d}+\phi x_{21}),
\end{flalign}
\end{subequations}
where the subscript $d$ signifies that the variable is being driven.  As
before,  $\mu$ represents a constant birth rate, $\beta$ is the contact rate,
$\phi$ is a constant that determines the antibody-dependent enhancement (ADE),
and $\sigma$ is the rate of recovery, with parameter values shown in Table~\ref{tab:parameters}.

Since $r_{1d}$ and $r_{2d}$ are decoupled from Eqs.~(\ref{e:s})-(\ref{e:xd2}),
the center manifold analysis of the following section will be performed using
the simpler 10-dimensional system given by Eqs.~(\ref{e:s})-(\ref{e:xd2}).

\section{Center Manifold Analysis}\label{cm}

We will reduce the dimension of the system given by
Eqs.~(\ref{e:s})-(\ref{e:xd2}) using the center manifold of the system.  The
analysis begins by determining the endemic equilibrium state of the system.
It is given as
\begin{equation}
(s_0,x_{i,0},r_{i,0},x_{ij,0},s_{d,0},x_{id,0})=
\left (\frac{\sigma}{\beta(1+\phi)},\frac{\mu}{2\sigma},\frac{\sigma}{\beta(1+\phi)},\frac{\mu}{2\sigma},\frac{\sigma}{\beta(1+\phi)},\frac{\mu}{2\sigma}\right )\label{e:ss}
\end{equation}
for all $i,j$.

A general nonlinear system may be transformed so that the system's
linear part has a block diagonal form consisting of three matrix blocks.  The
first matrix block will possess eigenvalues with positive real part; the
second matrix block will possess eigenvalues with negative real part; and the
third matrix block will possess eigenvalues with zero real part.  These three
matrix blocks are respectively associated with the unstable eigenspace, the
stable eigenspace, and the center eigenspace.  If there are no eigenvalues
with positive real part, then the orbits will rapidly decay to the center
eigenspace.

Equations~(\ref{e:s})-(\ref{e:xd2}) cannot be written in a block
diagonal form with one matrix block possessing eigenvalues with negative real
part and the other matrix block possessing eigenvalues with zero real part.
{Even though it is possible to construct a center manifold from a system
  not in separated block form~\citep{chilat97}, it is much easier to apply the center
  manifold theory to a system with separated stable and center directions.}
Therefore, we transform the original system given by
Eqs.~(\ref{e:s})-(\ref{e:xd2}) to a new system of equations that will have
the eigenvalue structure that is needed to apply center manifold theory.  The
theory allows one to find an invariant center manifold that passes through a
fixed point and to which one can restrict the new transformed system.

\subsection{Transformation of the Two-Serotype Model}\label{transformation}
To ease the analysis, we define a new set of variables, $\bar{s}$,
$\bar{x}_i$, $\bar{r}_i$, $\bar{x}_{ij}$, $\bar{s}_d$, and $\bar{x}_{id}$ for
all $i,j$ as
$\bar{s}(t)=s(t)-s_0$, $\bar{x}_i(t)=x_i(t)-x_{i,0}$,
$\bar{r}_i(t)=r(t)-r_{i,0}$, $\bar{x}_{ij}(t)=x_{ij}(t)-x_{ij,0}$,
$\bar{s}_d(t)=s_d(t)-s_{d,0}$, $\bar{x}_{id}(t)=x_{id}(t)-x_{id,0}$.
These new variables are substituted into Eqs.~(\ref{e:s})-(\ref{e:xd2}).

Then, treating $\mu$ as a small parameter, we rescale time by letting $t=\mu\tau$.
We may then introduce the following rescaled parameters:  $\beta=\beta_0/\mu$ and $\sigma=\sigma_0/\mu$,
where $\beta_0$ and $\sigma_0$ are $\mathcal{O}(1)$.  The inclusion of the
parameter $\mu$ as a new state variable means that the terms in our rescaled
system which contain $\mu$ are now nonlinear terms.  Furthermore, the system
is augmented with the auxiliary equation $\frac{d\mu}{d\tau}=0$.  The addition
of this auxiliary equation contributes an extra simple zero eigenvalue to the
system and adds one new center direction that has trivial dynamics.  The shifted
and rescaled, augmented system of equations now have the endemic fixed
point located at the origin and are provided in Appendix~\ref{sec:sra-sys}.

The Jacobian of these shifted and rescaled equations (Eqs.~(\ref{e:dsbardtau})-(\ref{e:dmudtau})) is computed to
zeroth-order in $\mu$ and is evaluated
at the origin. Ignoring the $\mu$ components, the Jacobian has only eight
linearly independent eigenvectors. Therefore, the
Jacobian is not diagonalizable.  However, it is possible to
transform Eqs.~(\ref{e:dsbardtau})-(\ref{e:dx2dbardtau}) to a block diagonal
form with {a separated} eigenvalue structure.  {As mentioned
  previously, this block structure makes the center manifold
analysis easier.}  We use a transformation matrix, ${\bf P}$, consisting of the
eight linearly independent eigenvectors of the Jacobian along with two other
vectors chosen to be linearly independent.  There are many choices for these
ninth and tenth vectors; our choice is predicated on keeping the vectors as simple as
possible.  This transformation matrix is given as follows:
\begin{equation}
\setlength{\arraycolsep}{4pt}
  \renewcommand{\arraystretch}{0.8}
{\bf P} = \left [ \begin{array}{cccccccccc}
0 & 0 & 0  & 0 & 1 & 0 & 0 & 0 & 0 & 0\\
-\phi & 0 & 0 & 0 & 0 & 1 & 0 & 0 & 0 & 0\\
0 & -\phi & 0 & 0 & 0 & 0 & 1 & 0 & 0 & 0\\
\phi& 0 & 0 & 0 & 0 & 0 & 0 & 1 & 0 & 0\\
0 & \phi & 0 & 0 & 0 & 0 & 0 & 0 & 1 & 0\\
1 & 0 & 0 & 0 & 0 & 1 & 0 & 0 & 0 & 0 \\
0 & 1 & 0 & 0 & 0 & 0 & 1 & 0 & 0 & 0\\
0 & 0 & 1 & 0 & 0 & 0 & 0 & 0 & 0 & 1\\
-\phi & 0 & \phi & -1 & 0 & 1 & 0 & 0 & 0 & 0 \\
0 & -\phi & 0 & 1 & 0 & 0 & 1 & 0 & 0 & 0 \\
 \end{array} \right ].\label{e:P}
\end{equation}

Using the fact that
$(\bar{s},\bar{x}_1,\bar{x}_2,\bar{r}_1,\bar{r}_2,\bar{x}_{21},\bar{x}_{12},\bar{s}_d,\bar{x}_{1d},\bar{x}_{2d})^T
= {\bf P}\cdot {\bf W}^T$, where ${\bf W}=(W_1,W_2,W_3,W_4,W_5,W_6,W_7,W_8,W_9,W_{10})$,
 then the
transformation matrix leads to the definition of new variables,
$W_i$, $i=1\ldots 10$ that can be found in Appendix~\ref{sec:def_new_var}.
The application of the transformation matrix to
Eqs.~(\ref{e:dsbardtau})-(\ref{e:dx2dbardtau}) leads to a set of
transformed evolution equations that are found in Appendix~\ref{sec:trans_evol_eq}.

\subsection{Application of the Center Manifold Theory}
The Jacobian of Eqs.~(\ref{e:dW1})-(\ref{e:dmu}) to zeroth-order in $\mu$ and
evaluated at the origin is
\begin{equation}
\setlength{\arraycolsep}{4pt}
  \renewcommand{\arraystretch}{0.8}
\left [ \begin{array}{cccc|cccccc}
-\sigma_0 & 0 & 0  & 0 & 0 & 0 & 0 & 0 & 0 & 0\\
0 & -\sigma_0 & 0 & 0 & 0 & 0 & 0 & 0 & 0 & 0\\
0 & 0 & -\frac{\phi\sigma_0}{1+\phi} & 0 & 0 & 0 & 0 & 0 & 0 & 0\\
0& 0 & 0 & -\frac{\phi\sigma_0}{1+\phi} & 0 & 0 & 0 & 0 & 0 & 0\\
\hline
0 & 0 & 0 & 0 & 0 & -\sigma_0 & -\sigma_0 & 0 & 0 & 0\\
0 & 0 & 0 & 0 & 0 & 0 & 0 & 0 & 0 & 0 \\
0 & 0 & 0 & 0 & 0 & 0 & 0 & 0 & 0 & 0\\
0 & 0 & 0 & 0 & 0 & \sigma_0 & -\sigma_0 & 0 & 0 & 0\\
0 & 0 & 0 & 0 & 0 & -\sigma_0 & \sigma_0 & 0 & 0 & 0 \\
0 & 0 & 0 & 0 & 0 & -\sigma_0 & -\sigma_0 & 0 & 0 & 0 \\
 \end{array} \right ],\label{e:jacobian}
\end{equation}
which shows that Eqs.~(\ref{e:dW1})-(\ref{e:dmu}) may be rewritten in the form
\begin{flalign}
&\frac{d{\bf x}}{d\tau}= {\bf A}{\bf x} + {\bf f}({\bf x},{\bf  y},\mu),\label{e:CMformx}\\
&\frac{d{\bf y}}{d\tau}= {\bf B}{\bf y} + {\bf g}({\bf x},{\bf  y},\mu),\label{e:CMformy}\\
&\frac{d\mu}{d\tau}=0,
\end{flalign}
where ${\bf x}=(W_1,W_2,W_3,W_4)$, ${\bf y}=(W_5,W_6,W_7,W_8,W_9,W_{10})$, ${\bf A}$ is a constant matrix with
eigenvalues that have negative real parts, ${\bf B}$ is a constant matrix with
eigenvalues that have zero real parts, and ${\bf f}$ and ${\bf g}$ are
nonlinear functions in ${\bf x}$, ${\bf y}$ and $\mu$.  In particular, ${\bf
  A}$ is the upper left block matrix of Eq.~(\ref{e:jacobian}), while ${\bf
  B}$ is the lower right block matrix of Eq.~(\ref{e:jacobian}).

Therefore, this new system of equations\b{,} which is an exact transformation of
Eqs.~(\ref{e:s})-(\ref{e:xd2})\b{,} will rapidly collapse onto a lower-dimensional manifold
 given by center manifold theory~\citep{car81,chilat97,dulusc03}. Furthermore,
 the lower dimensional center manifold is
 given by 
% \begin{subequations}
% \begin{flalign}
% W_1=&h_1\left (W_5,W_6,W_7,W_8,W_9,W_{10},\mu\right ),\label{e:h1}\\
% W_2=&h_2\left (W_5,W_6,W_7,W_8,W_9,W_{10},\mu\right ),\label{e:h2}\\
% W_3=&h_3\left (W_5,W_6,W_7,W_8,W_9,W_{10},\mu\right ),\label{e:h3}\\
% W_4=&h_4\left (W_5,W_6,W_7,W_8,W_9,W_{10},\mu\right ),\label{e:h4}
% \end{flalign}
% \end{subequations}
\begin{equation}
W_{i}=h_i(\bm y,\mu)\label{e:h1}
\end{equation}
where $h_i$, $i=1\ldots 4$ are unknown functions.

Substitution of the center manifold functions $W_i=h_i$ given by Eq.~(\ref{e:h1})
into the transformed evolution equations given in Appendix~\ref{sec:trans_evol_eq}  % Eq.~(\ref{e:h1}) into Eq.~(\ref{e:dW1}),  Eq.~(\ref{e:h2})
% into Eq.~(\ref{e:dW2}),  Eq.~(\ref{e:h3}) into Eq.~(\ref{e:dW3}), and
% Eq.~(\ref{e:h4}) into Eq.~(\ref{e:dW4})
leads to the center manifold condition
given in Appendix~\ref{sec:CM-condition}.

In general, it is not possible to solve the center manifold condition for the
four unknown functions $h_i(\bm y,\mu)$,
$i=1\ldots 4$. Therefore, a Taylor series expansion of
$h_i(\bm y,\mu)$, $i=1\ldots 4$ in {$\bm y$} and $\mu$ is
substituted into the four equations that comprise the center manifold
condition (Eqs.~(\ref{e:cmc-1})-(\ref{e:cmc-4})).  The unknown coefficients are
determined by equating terms of the same order, and the center manifold
equations are found to be
\begin{subequations}
\begin{flalign}
W_1 =& \frac{-\beta_0}{\sigma_0}W_5W_6 - 2\frac{\beta_0}{\sigma_0}W_6W_7   + \frac{\beta_0}{\sigma_0}W_6W_9+ \mathcal{O}(\epsilon^3),\label{e:W1cm}\\
W_2 =& \frac{-\beta_0}{\sigma_0}W_5W_7 - 2\frac{\beta_0}{\sigma_0}W_6W_7 + \frac{\beta_0}{\sigma_0}W_7W_8+ \mathcal{O}(\epsilon^3),\label{e:W2cm}\\
W_3 =& -\frac{\beta_0(1+\phi)^2}{\phi^2\sigma_0}W_5W_6 -
\frac{\beta_0(1+\phi)^2}{\phi^2\sigma_0}W_5W_7 +
\frac{\beta_0(1+\phi)^2}{\phi^2\sigma_0}W_6W_{10}  \nonumber\\
&+\frac{\beta_0(1+\phi)^2}{\phi^2\sigma_0}W_7W_{10}+ \mathcal{O}(\epsilon^3),\label{e:W3cm}\\
W_4 =& -\frac{\beta_0(1+\phi)^2}{\phi\sigma_0}W_5W_7 +
\frac{\beta_0(1+\phi)^2}{\phi\sigma_0}W_7W_{10}+ \mathcal{O}(\epsilon^3),\label{e:W4cm}
\end{flalign}
\end{subequations}
where $\epsilon =|(W_5,W_6,W_7,W_8,W_9,W_{10},\mu)|$ so that $\epsilon$ provides a count of the
number of $W_5$, $W_6$, $W_7$,
$W_8$, $W_9$, $W_{10}$, and $\mu$ factors in any one term. It is worth noting
that the center manifold equations may equivalently be found using a normal
form coordinate transform. Details of the method can be found in~\cite{rob08} and
application to epidemic models can be found in~\cite{fobisc09} and~\cite{Forgoston13}

Using the relations between $W_i$ variables and the ``barred'' variables given
by Eq.~(\ref{e:W1}),
Eqs.~(\ref{e:W1cm})-(\ref{e:W4cm}) can be equivalently written as follows:
\begin{subequations}
\begin{flalign}
&\sigma_0(\bar{x}_1-\bar{x}_{21})=
\beta_0(\bar{x}_1+\phi\bar{x}_{21})\left
  [\bar{s}-\bar{r}_2+\frac{3\phi}{(1+\phi)}\bar{x}_{12}+\frac{2-\phi}{(1+\phi)}\bar{x}_2\right ],\label{e:CM1}\\
&\sigma_0(\bar{x}_2-\bar{x}_{12})=\beta_0(\bar{x}_2+\phi\bar{x}_{12})\left
  [\bar{s}-\bar{r}_1+\frac{3\phi}{(1+\phi)}\bar{x}_{21}+\frac{2-\phi}{(1+\phi)}\bar{x}_1\right ],\label{e:CM2}\\
&\sigma_0\left [\bar{x}_{d1}+\bar{x}_{d2}-\bar{x}_1-\bar{x}_2\right
] = \frac{\beta_0(1+\phi)^2}{\phi^2}\left
  [\bar{x}_1+\bar{x}_2-\bar{x}_{d1}-\bar{x}_{d2}-\phi\bar{s}+\phi\bar{s}_d\right ]\times \nonumber\\
&\hspace{4cm}\left [ \frac{\bar{x}_1+\bar{x}_2+\phi\bar{x}_{12}+\phi\bar{x}_{21}}{(1+\phi)}\right ],\label{e:CM3}\\
&\sigma_0\left [\bar{x}_{d2}-\bar{x}_2\right
] = \frac{\beta_0(1+\phi)^2}{\phi^2}\left
  [\bar{x}_1+\bar{x}_2-\bar{x}_{d1}-\bar{x}_{d2}-\phi\bar{s}+\phi\bar{s}_d\right ]\times\left [ \frac{\bar{x}_2+\phi\bar{x}_{12}}{(1+\phi)}\right ].\label{e:CM4}
\end{flalign}
\end{subequations}

It is possible to simplify Eq.~(\ref{e:CM3}) by substituting Eq.~(\ref{e:CM4})
into Eq.~(\ref{e:CM3}).  The resulting simplified Eq.~(\ref{e:CM3}) is
\begin{equation}
\sigma_0\left [\bar{x}_{d1}-\bar{x}_1\right
] = \frac{\beta_0(1+\phi)^2}{\phi^2}\left
  [\bar{x}_1+\bar{x}_2-\bar{x}_{d1}-\bar{x}_{d2}-\phi\bar{s}+\phi\bar{s}_d\right ]\times\left [ \frac{\bar{x}_1+\phi\bar{x}_{21}}{(1+\phi)}\right ].\label{e:CM5}
\end{equation}

Solving Eqs.~(\ref{e:CM1})-(\ref{e:CM2}) for $\bar{x}_1$,
$\bar{x}_2$ leads to the following
approximation for the invariant manifold onto which the driver system collapses:
\begin{subequations}
\begin{flalign}
&\bar{x}_1 = \frac{\sigma_0\bar{x}_{21}+\beta_0\phi\bar{x}_{21}\left [
    \bar{s}-\bar{r}_2+\frac{2-\phi}{1+\phi}\bar{x}_2+\frac{3\phi}{1+\phi}\bar{x}_{12}\right ]}{\sigma_0-\beta_0\left [ \bar{s}-\bar{r}_2+\frac{2-\phi}{1+\phi}\bar{x}_2+\frac{3\phi}{1+\phi}\bar{x}_{12}\right ]}\label{e:CM1-2}\\
&\bar{x}_2 = \frac{\sigma_0\bar{x}_{12}+\beta_0\phi\bar{x}_{12}\left [
    \bar{s}-\bar{r}_1+\frac{2-\phi}{1+\phi}\bar{x}_1+\frac{3\phi}{1+\phi}\bar{x}_{21}\right ]}{\sigma_0-\beta_0\left [ \bar{s}-\bar{r}_1+\frac{2-\phi}{1+\phi}\bar{x}_1+\frac{3\phi}{1+\phi}\bar{x}_{21}\right ]}.\label{e:CM2-2}
%&\bar{x}_{d1} = \frac{\sigma_0\bar{x}_1+\frac{\beta_0\left( 1+\phi\right )^2}{\phi^2} \left [
%    \bar{x}_1+\bar{x}_2-\bar{x}_{d2}-\phi\bar{s}+\phi\bar{s}_d\right ]\left [
% \frac{\bar{x}_1+\phi\bar{x}_{21}}{1+\phi} \right
%]}{\sigma_0+\frac{\beta_0\left (1+\phi\right )^2}{\phi^2}\left [
%  \frac{\bar{x}_1+\phi\bar{x}_{21}}{1+\phi}\right ]},\label{e:CM3-2}\\
%&\bar{x}_{d2} = \frac{\sigma_0\bar{x}_2+\frac{\beta_0\left( 1+\phi\right )^2}{\phi^2} \left [
%    \bar{x}_1+\bar{x}_2-\bar{x}_{d1}-\phi\bar{s}+\phi\bar{s}_d\right ]\left [
% \frac{\bar{x}_2+\phi\bar{x}_{12}}{1+\phi} \right
%]}{\sigma_0+\frac{\beta_0\left (1+\phi\right )^2}{\phi^2}\left [ \frac{\bar{x}_2+\phi\bar{x}_{12}}{1+\phi}\right ]}.\label{e:CM4-2}
\end{flalign}
\end{subequations}
This invariant manifold was found previously in \cite{shbisc07} and represents the close relationship of primary infectives and secondary infectives.

Similarly, solving the simplified Eq.~(\ref{e:CM3})
given by Eq.~(\ref{e:CM5}) and Eqs.~(\ref{e:CM4}) simultaneously for $\bar{x}_{d1}$ and $\bar{x}_{d2}$ leads to an approximate invariant manifold for the driven primary infectives expressed almost entirely in terms of driver system variables:
\begin{subequations}
\begin{flalign}
&\bar{x}_{d1}=\frac{\left (\sigma_0 \bar{x}_1\left (1+\phi\right )
+\Gamma\left [ \bar{x}_1\left
    (\bar{x}_1+\bar{x}_2 +\phi\left (\bar{x}_{12}+\bar{x}_{21}\right )\right )
  + \phi\left (\bar{s}_d-\bar{s}\right )\left
    (\bar{x}_1+\phi\bar{x}_{21}\right )\right ]\right )}{\sigma_0\left
(1+\phi\right ) +\Gamma\left (\bar{x}_1+\bar{x}_2+\phi\left
  (\bar{x}_{12}+\bar{x}_{21}\right )\right )}\label{e:CM3-2}\\
&\bar{x}_{d2}=\frac{\left (\sigma_0 \bar{x}_2\left (1+\phi\right )
+\Gamma\left [ \bar{x}_2\left
    (\bar{x}_1+\bar{x}_2 +\phi\left (\bar{x}_{12}+\bar{x}_{21}\right )\right )
  + \phi\left (\bar{s}_d-\bar{s}\right )\left
    (\bar{x}_2+\phi\bar{x}_{12}\right )\right ]\right )}{\sigma_0\left
(1+\phi\right ) +\Gamma\left (\bar{x}_1+\bar{x}_2+\phi\left
  (\bar{x}_{12}+\bar{x}_{21}\right )\right )}\label{e:CM4-2}
\end{flalign}
\end{subequations}
where $\Gamma=\frac{\beta_0\left (1+\phi\right )^2}{\phi^2}$.  Although these equations do not prove that the driven system synchronizes to the driver system, they do indicate that the driven system rapidly approaches a manifold in which the driven primary infectives are strongly affected by the infectives in the driver system.  We show in subsequent numerical simulations that the driven variables do in fact approach those in the driver system.

\section{Results} \label{results}
\subsection{Comparison of Solutions}
The original equations in the shifted, barred variables (Eqs.~(\ref{e:dsbardtau})-(\ref{e:dx2dbardtau})) are solved, and the
values of the variables found on the right hand side of
Eqs.~(\ref{e:CM1-2})-(\ref{e:CM4-2}) are substituted into
Eqs.~(\ref{e:CM1-2})-(\ref{e:CM4-2}) to find the value of $\bar{x}_1$,
$\bar{x}_2$, $\bar{x}_{d1}$, and $\bar{x}_{d2}$ given by the center manifold
equations.  After computing in the shifted variables, we shift back to the
original variables.

\begin{figure}[b!]
\begin{center}
\begin{minipage}{0.49\linewidth}
\includegraphics[width=6.5cm]{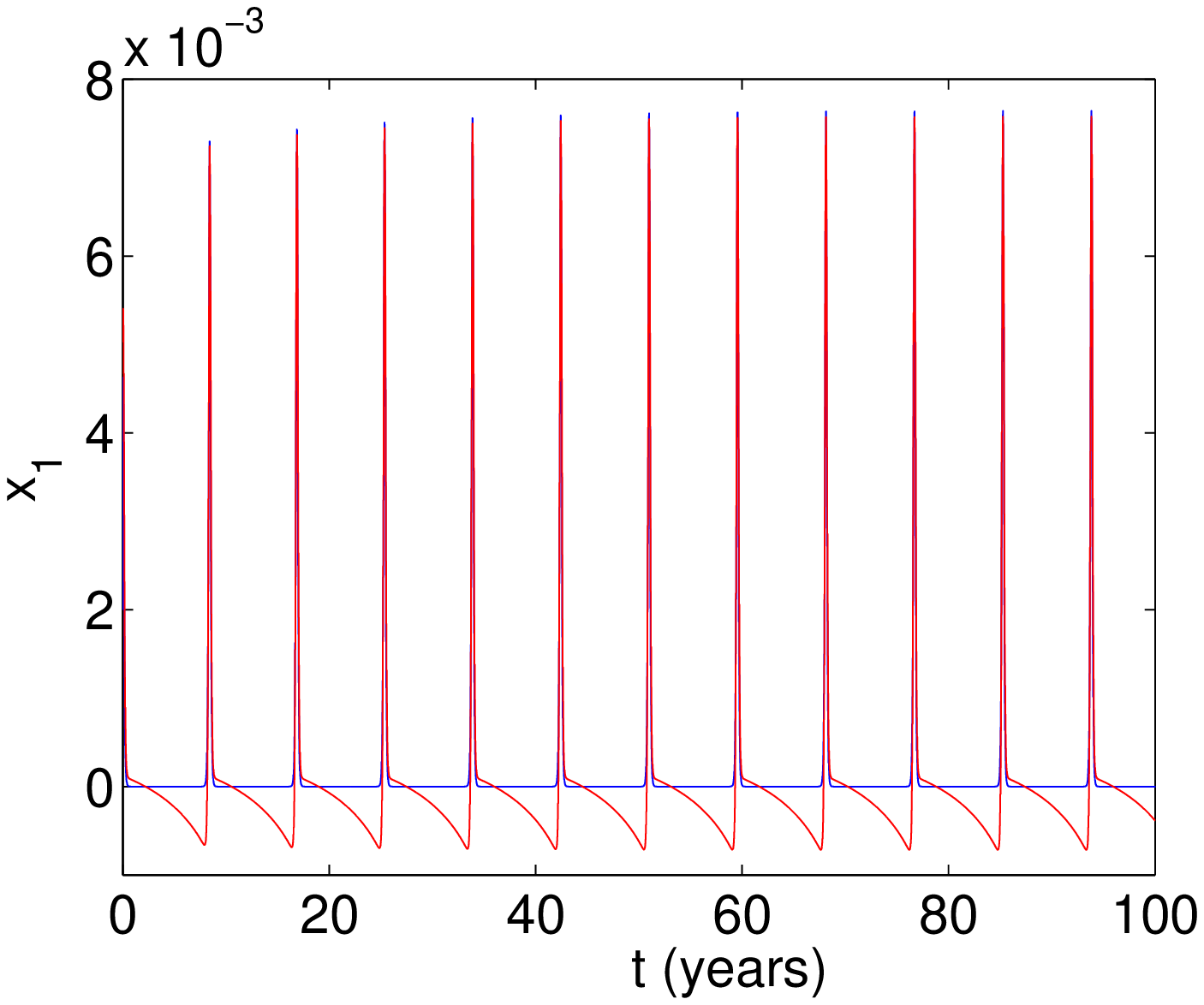}
\end{minipage}
\begin{minipage}{0.49\linewidth}
\includegraphics[width=6.5cm]{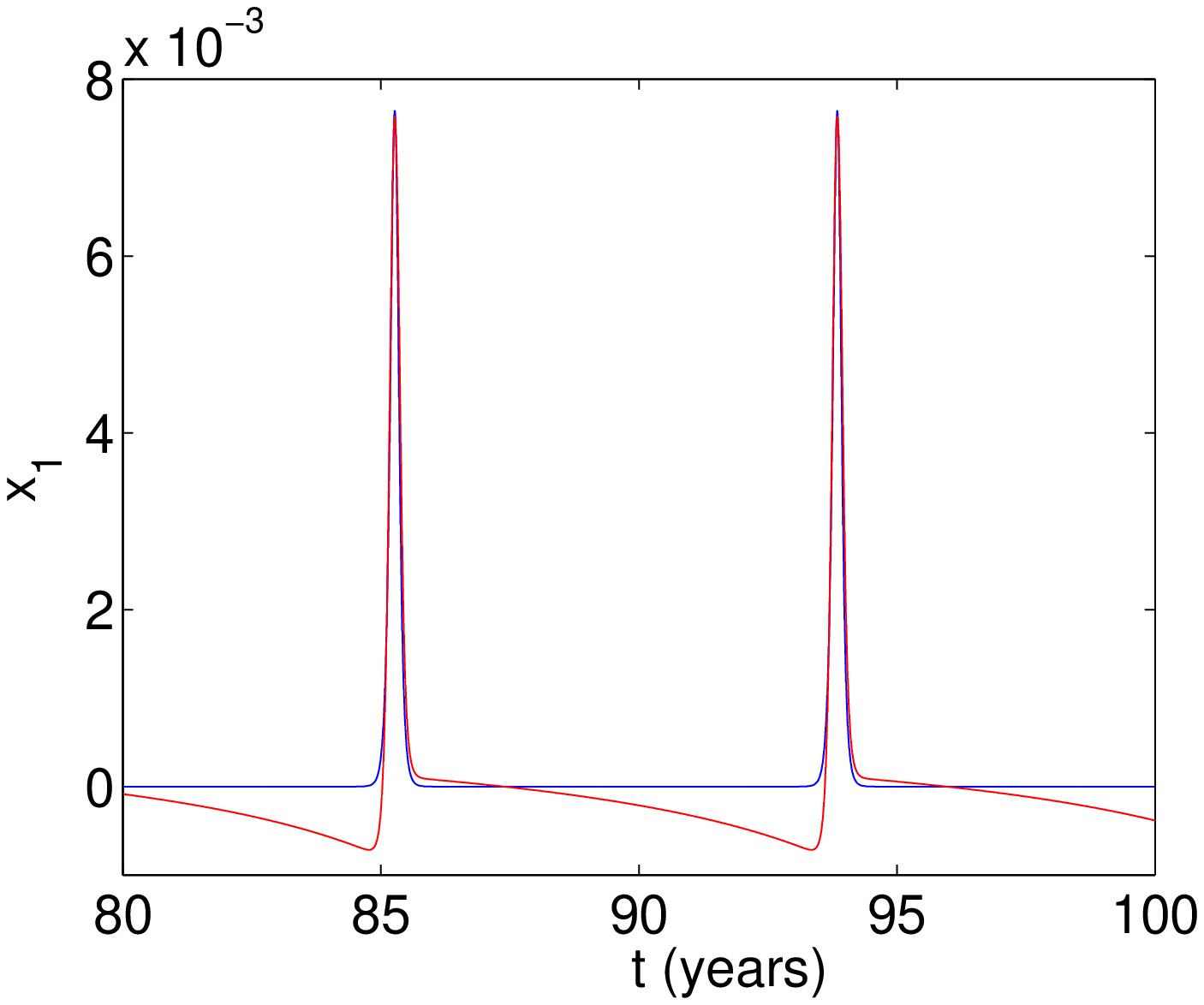}
\end{minipage}
\caption{\label{fig:x1compare}Plot of $x_1$ computed using the complete system
  of equations (blue) and the center manifold equation (red) for (a) $t=0$ to
  $t=100$, and (b) $t=80$ to $t=100$.  Parameter values are given by
  $\mu=0.02$, $\sigma=50.0$, $\phi=3.0$, and $\beta=200.0$.}
\end{center}
\end{figure}

Figure~\ref{fig:x1compare}(a) shows a comparison of $x_1$
found through computation of the complete system and found through computation
using the center manifold equation.  To show the agreement more clearly,
Fig.~\ref{fig:x1compare}(b) shows a portion of Fig.~\ref{fig:x1compare}(a). The
prediction for the primary infectives is occasionally negative since the center manifold
equations involve $\bar{x}_i$ terms which represent infectives shifted from
their fixed point. As discussed in~\cite{shbisc07},
the predicted deviations may be large enough so that adding the fixed point
$x_{i,0}$ results in a negative value for the $x_i$ terms. However, since the
errors occur in regions where there is no epidemic outbreak, they are not of
biological importance. Furthermore, one can see that the timing of the
outbreaks predicted by the reduced system agrees well with the actual outbreak time of the complete system.

\begin{figure}[h!]
\begin{center}
\begin{minipage}{0.49\linewidth}
\includegraphics[width=6.5cm]{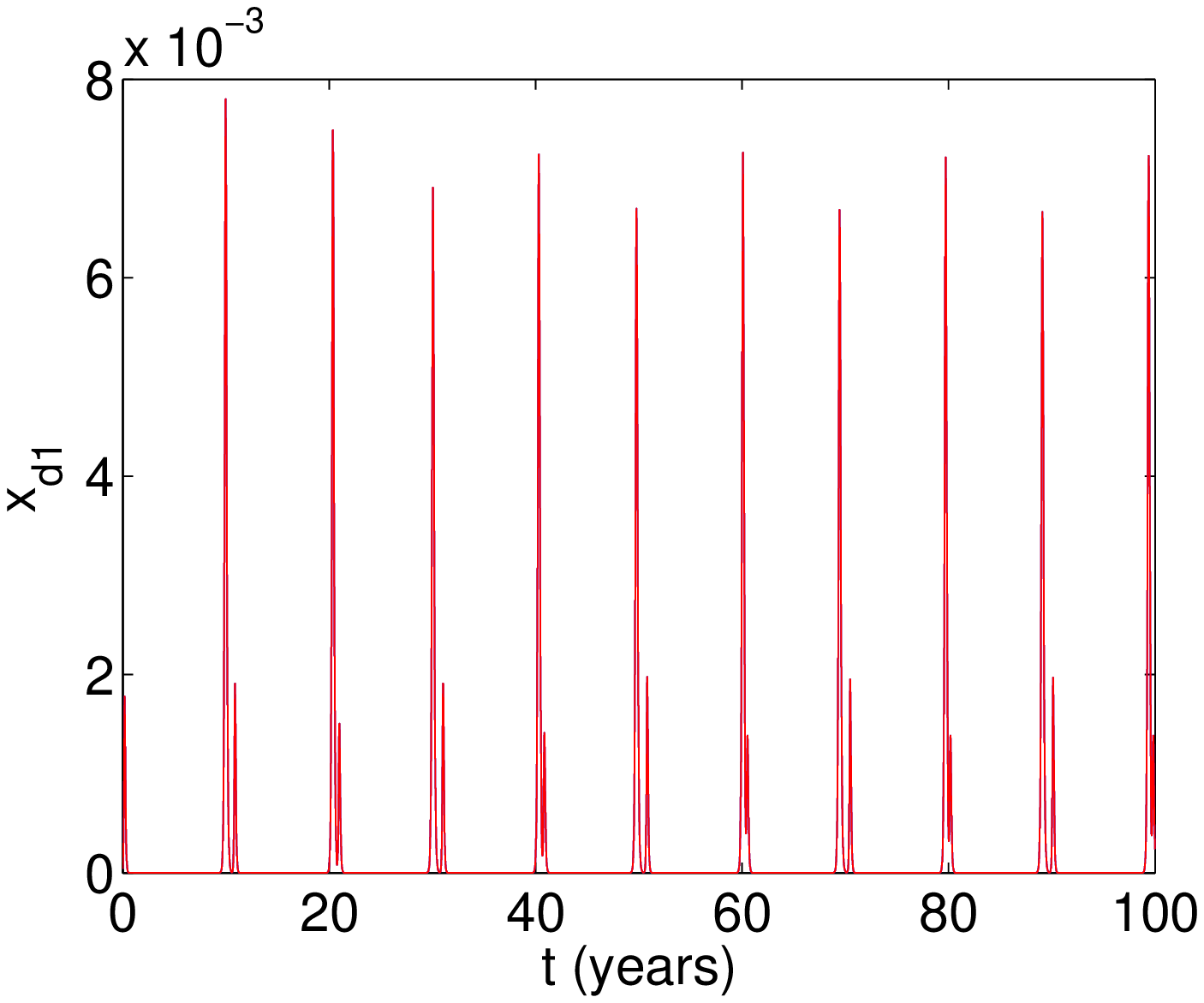}
\end{minipage}
\begin{minipage}{0.49\linewidth}
\includegraphics[width=6.5cm]{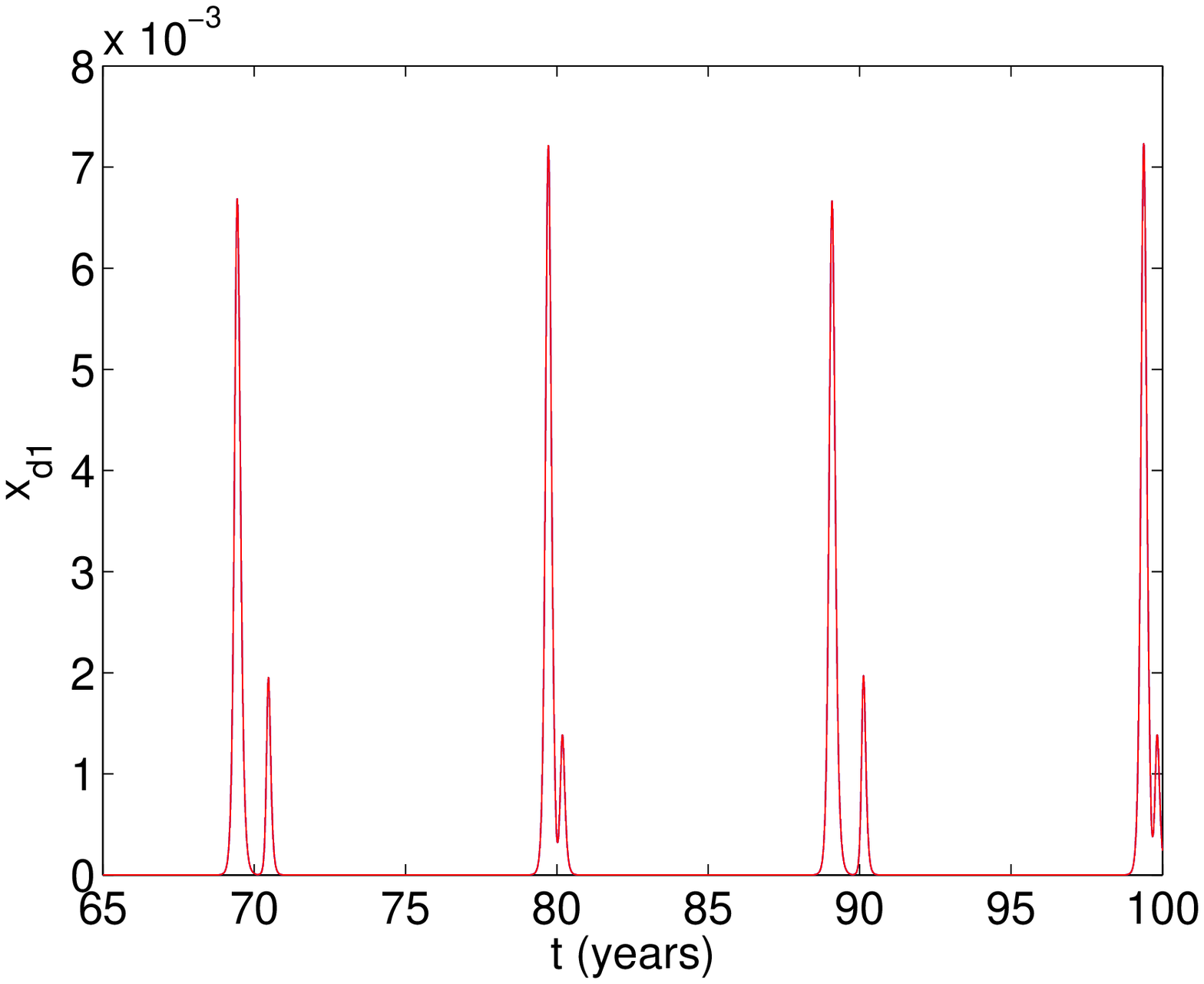}
\end{minipage}\\
\begin{minipage}{0.49\linewidth}
\includegraphics[width=6.5cm]{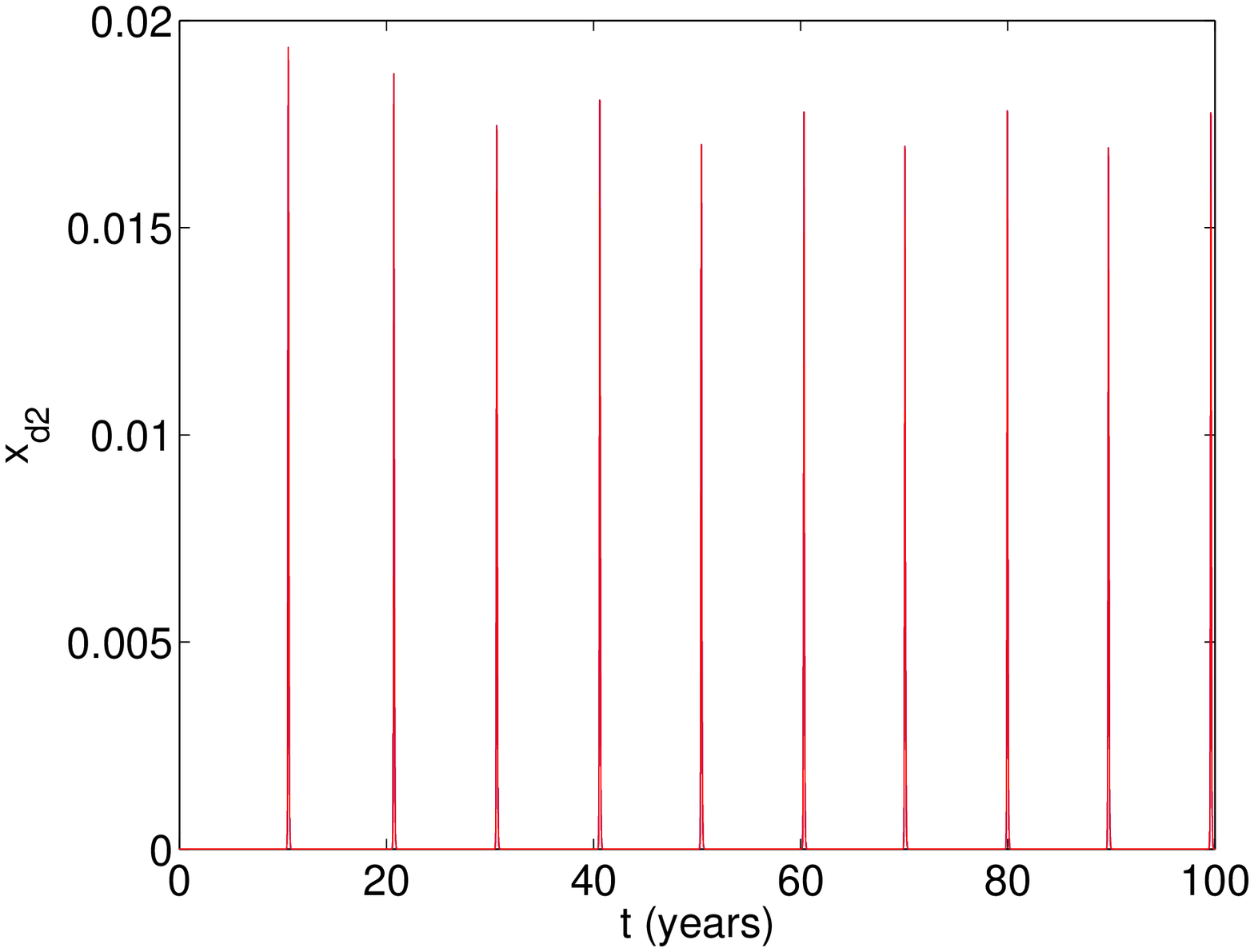}
\end{minipage}
\begin{minipage}{0.49\linewidth}
\includegraphics[width=6.5cm]{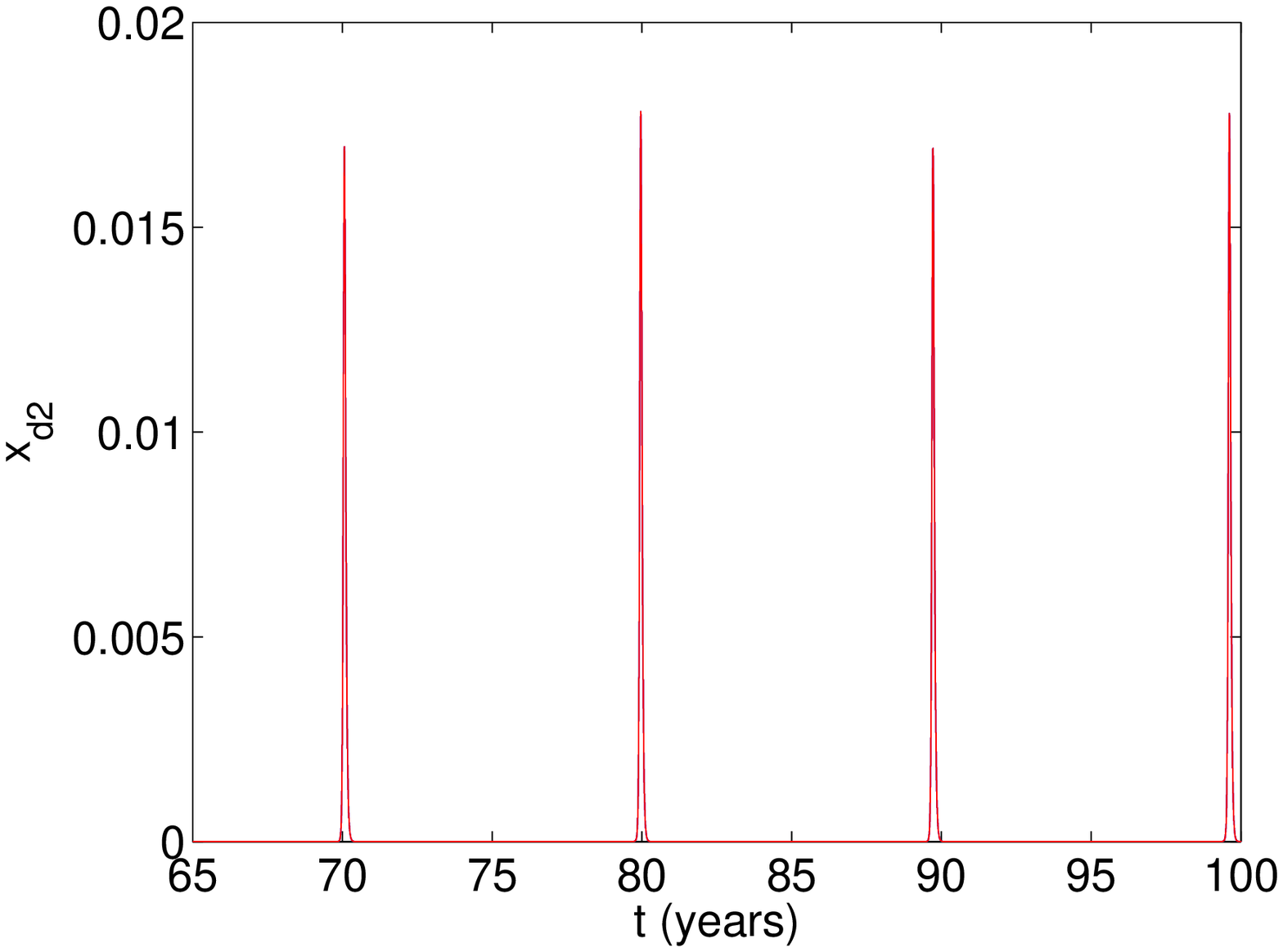}
\end{minipage}
\caption{\label{fig:xd1xd2compare}Plots of (a)-(b) $x_{d1}$ and (c)-(d) $x_{d2}$ computed using the complete
system of equations (blue) and the center manifold equation (red) for (a) and
(c) $t=0$ to
$t=100$, and (b) and (d) $t=65$ to $t=100$.  Parameter values are given by $\mu=0.02$, $\sigma=50.0$, $\phi=3.0$, and $\beta=200.0$.}
\end{center}
\end{figure}

Figure~\ref{fig:xd1xd2compare}(a) and (c) shows the comparison of $x_{d1}$ and
$x_{d2}$ respectively
found through computation of the complete system and found through computation
using the center manifold equation. The agreement is perfect and is seen more
clearly in Figs.~\ref{fig:xd1xd2compare}(b) and (d) which show portion of
Figs.~\ref{fig:xd1xd2compare}(a) and (c) respectively.
% \begin{figure}[h!]
% \begin{center}
% \includegraphics[width=12.5cm]{x2compare_newparam}
% \caption{\label{fig:x2compare}Plot of $x_2$ computed using the complete
% system of equations (blue) and the center manifold equation (red) for $t=0$ to
% $t=100$.  Parameter values are given by $\mu=0.02$, $\sigma=50.0$, $\phi=3.0$, and $\beta=200.0$.}
% \end{center}
% \end{figure}

%\begin{figure}[h!]
%\begin{center}
%\includegraphics[width=12.5cm]{xd1xd2compare_newparam}
%\caption{\label{fig:xd1xd2compare}Plot of $x_{d1}$ and $x_{d2}$ computed using the complete
%system of equations (blue) and the center manifold equation (red) for $t=0$ to
%$t=100$.  Parameter values are given by $\mu=0.02$, $\sigma=50.0$, $\phi=3.0$, and $\beta=200.0$.}
%\end{center}
%\end{figure}

% \begin{figure}[h!]
% \begin{center}
% \subfigure[]{\includegraphics[width=9.5cm]{xd1_052714}}
% \subfigure[]{\includegraphics[width=9.5cm]{xd2_052714}}
% \caption{\label{fig:xd1xd2compare}Plots of $x_{d1}$ and $x_{d2}$ computed using the complete
% system of equations (blue) and the center manifold equation (red) for $t=0$ to
% $t=100$.  Parameter values are given by $\mu=0.02$, $\sigma=50.0$, $\phi=3.0$, and $\beta=200.0$.}
% \end{center}
% \end{figure}

\subsection{Transverse Stability}

To show that the driven system synchronizes with the full system for the
components chosen, we can examine the transverse stability of the solution
of the subsystem. That is, we examine the difference between the full
vector field (the driver) and the subsystem. By computing the linear
variational equations of the difference,  we can find the conditional Lyapunov exponents
for solutions near the subsystem.

To establish notation, we rewrite the components of the governing equations for the two serotype
multistrain disease model as
\begin{equation}
{\bf X}=[s,x_{1},x_{2}]^T \quad {\bf Y}=[r_{1},r_{2},x_{21},x_{12}]^T,
\end{equation}
where $T$ denotes the transpose.
Additionally, we can rewrite the components of the governing equations for the two serotype
multistrain disease subsystem that are driven by the secondary infectious individuals as
\begin{equation}
{\bf Z}=[s_d,x_{1d},x_{2d}]^T.
\end{equation}

To write the system of equations we consider the split differential equations
\begin{subequations}
\begin{flalign}
&{\bf {\bf \dot{X}}=F_{1}(X,Y)} \\
&{\bf {\bf \dot{Y}}=F_{2}(X,Y).}
\end{flalign}
along with the subsystem
\begin{equation}
{\bf {\bf \dot{Z}}=F_{1}(Z,Y)}.
\end{equation}
\end{subequations}

\begin{figure}[h!]
\begin{center}
\includegraphics[width=12.5cm]{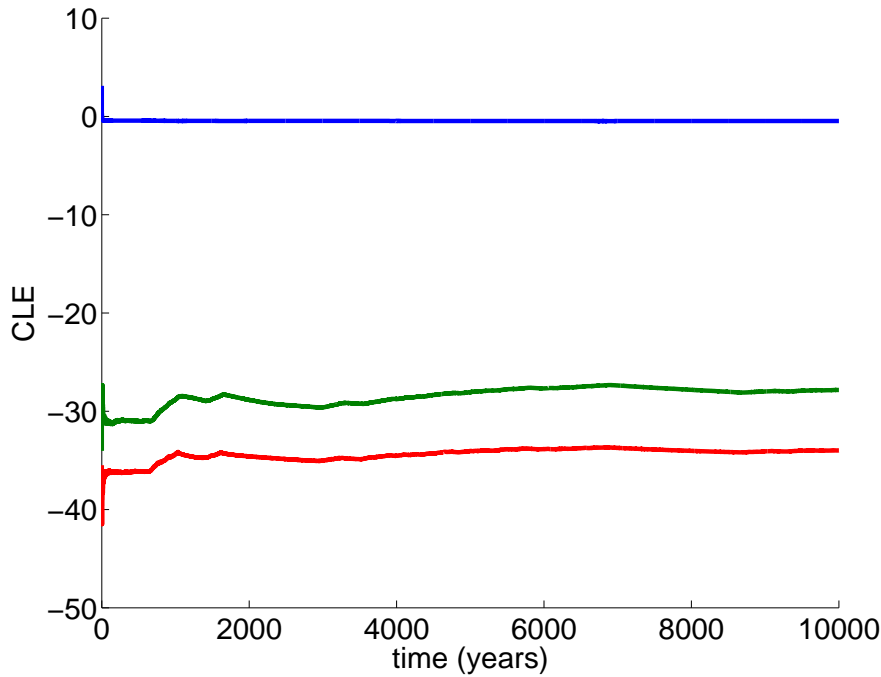}
\caption{\label{fig:CLE} Time series of the conditional Lyapunov exponents.
  Parameter values are given by $\mu=0.02$, $\sigma=50.0$, $\phi=3.0$, and
  $\beta=200.0$. The initial condition for the driver system was a point on
  the chaotic attractor obtained by computing a long time series, and the
  initial condition for the driven system was a perturbation from the driver.}
\end{center}
\end{figure}

We would like ${\bf Z}(t)\rightarrow{\bf X}(t)$ in the asymptotic limit.
Letting ${\bm \xi}={\bf X}-{\bf Z}$, one obtains the linear variation
about ${\bf X}(t)$ given by ${\dot{\bm \xi}}={\bf F}_{1,x}({\bf X},{\bf Y}){\bm \xi}$.
The Lyapunov exponents of the subsystem are the conditional Lyapunov
exponents (CLE). Negative CLE provide a sufficient condition
for the subsystem to converge, and are shown in Fig.~\ref{fig:CLE}.

% \subsection{Rates of convergence}
% \r{I am not sure if this section needs to be here as a separate entity. And do
% we want to show time series of complexity?}

% Relevant figures are Fig.~\ref{fig:errortimeseries} and Fig.~\ref{fig:synctime}.

\begin{figure}[h!]
\begin{center}
\includegraphics[width=12.5cm]{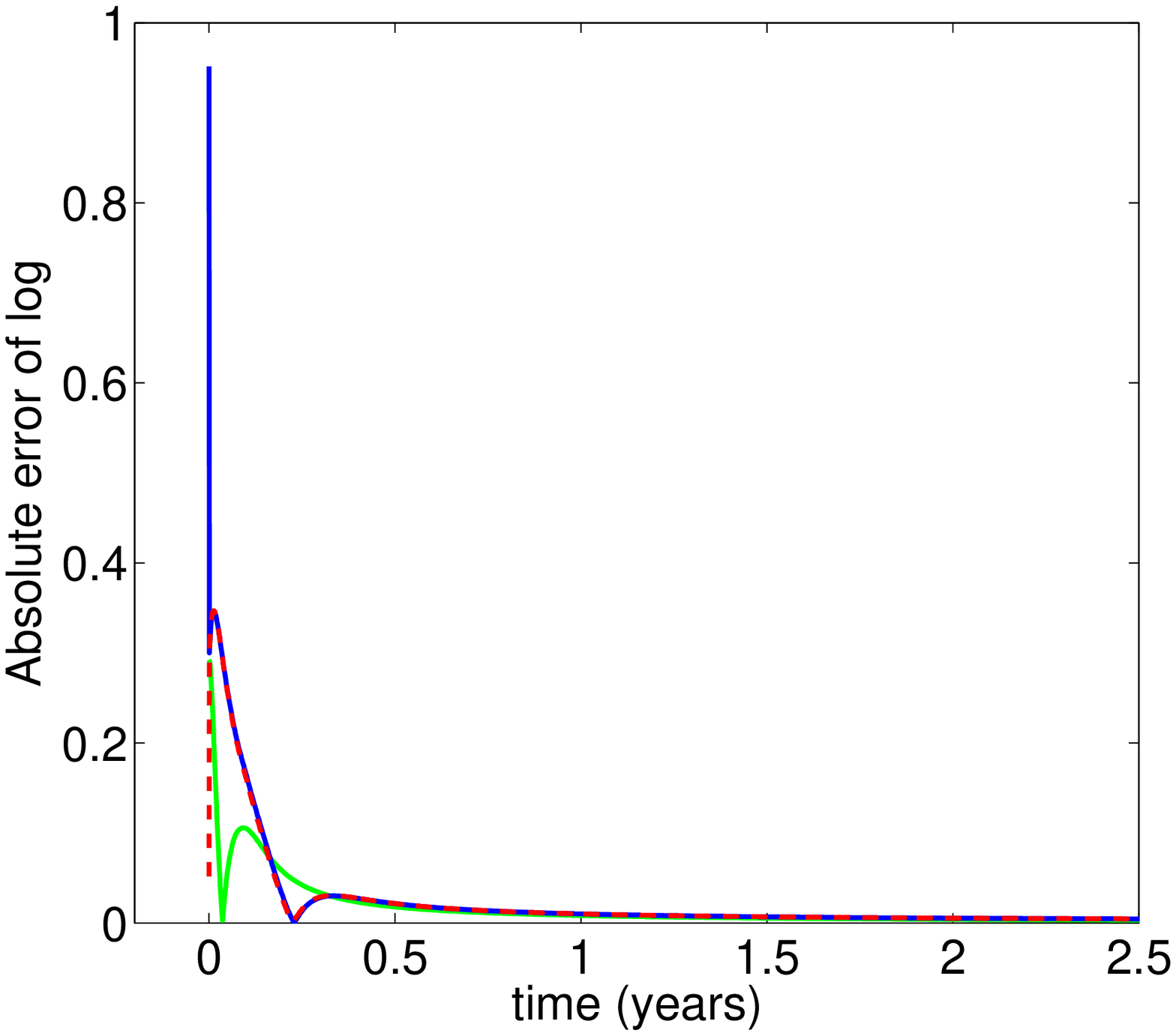}
\caption{Absolute error in log variables for susceptibles and primary
  infectives versus time.  Solid green:  $|\log s_d - \log s|$, solid blue:
  $|\log x_{1d}-\log x_1|$, dashed red:  $|\log x_{2d}-\log x_2|$.  The
  initial condition for the driver system was a point on the chaotic attractor
  obtained by computing a long time series. The initial condition for the driven system was a
  random perturbation with mean 10\% away from the driver system.  Parameter
  values are the same as in Fig.~\ref{fig:CLE}.} 
\label{fig:errortimeseries}
\end{center}
\end{figure}

In Fig.~\ref{fig:errortimeseries}, we show a sample time series of the error between the driven and driver systems.  Susceptibles and primary infectives in the driven system rapidly converge to the values found in the driver system.  Even when the driven system is initially perturbed farther from the driver system, convergence occurs within a similar time period.

% \begin{figure}[h!]
% \begin{center}
% \includegraphics[width=12.5cm]{Time_to_threshold_1Ksamples.eps}
% \caption{Synchronization time for various perturbations of the driven system.  Driven system initial conditions for each parameter value were 100 random initial conditions in a ball of radius $\epsilon$ of the driver initial condition.  System was considered to have attained synchronization when the absolute difference between driven and driver primary infectives \textbf{[which strain?]} was less than $5\times 10^{-9}$. Parameter values are ???  \textbf{[Maybe redo with a less stringent synchronization criterion.]}} \label{fig:synctime}
% \end{center}
% \end{figure}

\section{Conclusions} \label{sec:conc}

In this article, we considered a two serotype multistrain disease model in
which susceptibles may develop a primary infection from either of the two
serotypes and upon recovery become immune to the strain that caused the
primary infection.  However, the second serotype may cause a secondary
infection. We also consider a two serotype multistrain disease subsystem that
is driven by the secondary infectious individuals of the first model. By
performing center manifold analysis, we can reduce the dimension of these
driven and driver systems. In particular, we were able to analytically find an approximate
invariant manifold for the driven primary infectives expressed almost entirely
in terms of driver system variables.
Numerical simulations demonstrate the excellent agreement between solutions of
the original, higher-dimensional system and the lower-dimensional center
manifold equations. We further demonstrated the driven system is synchronized
with the complete system by examining the transverse stability of the
subsystem solution.

In summary, we have developed a new method to determine unobserved epidemic
sub-populations using synchronization properties of the epidemic model. As an
application, it is known that the majority of dengue fever hospital cases are
secondary infections \cite{nisa03}.  Thus one may have information about secondary infections
but not about primary infections. In driving a model using the known
secondary infection data, the primary infective populations in the driven
system synchronize to the correct values from the driver system. It is
therefore possible to deduce unobserved primary infection levels in the population.
It should be noted that this method probably works for other multistrain
models, such as the dengue model in \cite{Bianco09} in which we include temporary cross immunity immediately following a primary infection.

Future work includes the application of our method to real disease data.
Before this can be performed, one would first have to perform the analysis in
the presence of noise (i.e. the driver signal comes from a noisy system and/or
has additional errors in it).  Additionally, the theory will need to be
applied to the situation when data is sampled discretely.  Preliminary testing
using discretely sampled data for the secondary infectives has shown that the
technique can indeed recover primary infectives.

\begin{acknowledgements}
EF is supported by Award Number CMMI-1233397 from the National Science
  Foundation. LBS is supported by Award Number R01GM090204 from the National Institute of General Medical Sciences. The content is solely the
responsibility of the authors and does not necessarily represent the official
views of the National Institute Of General Medical Sciences or the National
Institutes of Health. IBS is supported by the NRL Base Research Program contract number N0001414WX00023 and by the Office of Naval Research contract number N0001414WX20610.
\end{acknowledgements}

\clearpage
%\bibliographystyle{bmb}
%\bibliography{Refs}

\clearpage

\appendix

\section{Shifted, Rescaled, and Augmented System of Equations}\label{sec:sra-sys}

The governing equations for the two serotype multistrain disease model and the
subsystem driven by secondary infectious individuals are given by Eqs.~(\ref{e:s})-(\ref{e:xd2}).
We define a new set of variables, $\bar{s}$,
$\bar{x}_i$, $\bar{r}_i$, $\bar{x}_{ij}$, $\bar{s}_d$, and $\bar{x}_{id}$ for
all $i,j$ as
$\bar{s}(t)=s(t)-s_0$, $\bar{x}_i(t)=x_i(t)-x_{i,0}$,
$\bar{r}_i(t)=r(t)-r_{i,0}$, $\bar{x}_{ij}(t)=x_{ij}(t)-x_{ij,0}$,
$\bar{s}_d(t)=s_d(t)-s_{d,0}$, $\bar{x}_{id}(t)=x_{id}(t)-x_{id,0}$,
and these new variables are substituted into Eqs.~(\ref{e:s})-(\ref{e:xd2}).

Then, treating $\mu$ as a small parameter, we rescale time by letting $t=\mu\tau$.
We may then introduce the following rescaled parameters:  $\beta=\beta_0/\mu$ and $\sigma=\sigma_0/\mu$,
where $\beta_0$ and $\sigma_0$ are $\mathcal{O}(1)$.  The inclusion of the
parameter $\mu$ as a new state variable means that the terms in our rescaled
system which contain $\mu$ are now nonlinear terms.  Furthermore, the system
is augmented with the auxiliary equation $\frac{d\mu}{d\tau}=0$.  The addition
of this auxiliary equation contributes an extra simple zero eigenvalue to the
system and adds one new center direction that has trivial dynamics.  The shifted
and rescaled, augmented system of equations is given as
\begin{subequations}
\begin{flalign}
&\frac{d\bar{s}}{d\tau} = \mu^2 - \beta_0 \left (
  \bar{s}+\frac{\sigma_0}{\beta_0 (1+\phi)}\right ) \left
  (\bar{x}_1+\bar{x}_2+\phi\left
    (\bar{x}_{21}+\bar{x}_{12}\right )+\frac{\mu^2(1+\phi)}{\sigma_0}\right
),\label{e:dsbardtau}\\
&\frac{d\bar{x}_1}{d\tau} = \beta_0\left ( \bar{s}+\frac{\sigma_0}{\beta_0
    (1+\phi)}\right )\left
  (\bar{x}_1+\phi\bar{x}_{21}+\frac{\mu^2(1+\phi)}{2\sigma_0}\right
)-\sigma_0\left (\bar{x}_1+\frac{\mu^2}{2\sigma_0}\right ),\\
&\frac{d\bar{x}_2}{d\tau} = \beta_0\left ( \bar{s}+\frac{\sigma_0}{\beta_0
    (1+\phi)}\right )\left
  (\bar{x}_2+\phi\bar{x}_{12}+\frac{\mu^2(1+\phi)}{2\sigma_0}\right
)-\sigma_0\left (\bar{x}_2+\frac{\mu^2}{2\sigma_0}\right ),\\
&\frac{d\bar{r}_1}{d\tau} =\sigma_0\left (\bar{x}_1+\frac{\mu^2}{2\sigma_0}\right ) -\beta_0\left ( \bar{r}_1+\frac{\sigma_0}{\beta_0
    (1+\phi)}\right )\left
  (\bar{x}_2+\phi\bar{x}_{12}+\frac{\mu^2(1+\phi)}{2\sigma_0}\right
),\\
&\frac{d\bar{r}_2}{d\tau} =\sigma_0\left (\bar{x}_2+\frac{\mu^2}{2\sigma_0}\right ) -\beta_0\left ( \bar{r}_2+\frac{\sigma_0}{\beta_0
    (1+\phi)}\right )\left
  (\bar{x}_1+\phi\bar{x}_{21}+\frac{\mu^2(1+\phi)}{2\sigma_0}\right
),\\
&\frac{d\bar{x}_{21}}{d\tau} = \beta_0\left ( \bar{r}_2+\frac{\sigma_0}{\beta_0
    (1+\phi)}\right )\left
  (\bar{x}_1+\phi\bar{x}_{21}+\frac{\mu^2(1+\phi)}{2\sigma_0}\right
)-\sigma_0\left (\bar{x}_{21}+\frac{\mu^2}{2\sigma_0}\right ),\\
&\frac{d\bar{x}_{12}}{d\tau} = \beta_0\left ( \bar{r}_1+\frac{\sigma_0}{\beta_0
    (1+\phi)}\right )\left
  (\bar{x}_2+\phi\bar{x}_{12}+\frac{\mu^2(1+\phi)}{2\sigma_0}\right
)-\sigma_0\left (\bar{x}_{12}+\frac{\mu^2}{2\sigma_0}\right ),\\
&\frac{d\bar{s}_d}{d\tau} = \mu^2 - \beta_0 \left (
  \bar{s}_d+\frac{\sigma_0}{\beta_0 (1+\phi)}\right ) \left
  (\bar{x}_{1d}+\bar{x}_{2d}+\phi\left
    (\bar{x}_{21}+\bar{x}_{12}\right )+\frac{\mu^2(1+\phi)}{\sigma_0}\right
),\\
&\frac{d\bar{x}_{1d}}{d\tau} = \beta_0\left ( \bar{s}_d+\frac{\sigma_0}{\beta_0
    (1+\phi)}\right )\left
  (\bar{x}_{1d}+\phi\bar{x}_{21}+\frac{\mu^2(1+\phi)}{2\sigma_0}\right
)-\sigma_0\left (\bar{x}_{1d}+\frac{\mu^2}{2\sigma_0}\right ),\\
&\frac{d\bar{x}_{2d}}{d\tau} = \beta_0\left ( \bar{s}_d+\frac{\sigma_0}{\beta_0
    (1+\phi)}\right )\left
  (\bar{x}_{2d}+\phi\bar{x}_{12}+\frac{\mu^2(1+\phi)}{2\sigma_0}\right
)-\sigma_0\left (\bar{x}_{2d}+\frac{\mu^2}{2\sigma_0}\right ),\label{e:dx2dbardtau}\\
&\frac{d\mu}{d\tau}=0,\label{e:dmudtau}
\end{flalign}
\end{subequations}
where the endemic fixed
point is now located at the origin.

\clearpage

\section{Definition of New Variables}\label{sec:def_new_var}
% \begin{subequations}
% \begin{flalign}
% W_1 =& \frac{\bar{x}_{21}-\bar{x}_1}{1+\phi},\label{e:W1}\\
% W_2 =& \frac{\bar{x}_{12}-\bar{x}_2}{1+\phi},\label{e:W2}\\
% W_3 =& \frac{\bar{x}_{1d}+\bar{x}_{2d}-\bar{x}_1-\bar{x}_2}{\phi},\label{e:W3}\\
% W_4 =& \bar{x}_{2d}-\bar{x}_2,\label{e:W4}\\
% W_5 =& \bar{s},\label{e:W5}\\
% W_6 =& \frac{\bar{x}_{1}+\phi\bar{x}_{21}}{1+\phi},\label{e:W6}\\
% W_7 =& \frac{\bar{x}_{2}+\phi\bar{x}_{12}}{1+\phi},\label{e:W7}\\
% W_8 =&
% \frac{\phi\bar{r}_{1}+\phi\bar{x}_{1}+\bar{r}_1-\phi\bar{x}_{21}}{1+\phi},\label{e:W8}\\
% W_9 =&
% \frac{\phi\bar{r}_{2}+\phi\bar{x}_{2}+\bar{r}_2-\phi\bar{x}_{12}}{1+\phi},\label{e:W9}\\
% W_{10} =& \frac{\phi\bar{s}_{d}-\bar{x}_{1d}-\bar{x}_{2d}+\bar{x}_{1}+\bar{x}_2}{\phi}.\label{e:W10}
% \end{flalign}
% \end{subequations}

Using the fact that
$(\bar{s},\bar{x}_1,\bar{x}_2,\bar{r}_1,\bar{r}_2,\bar{x}_{21},\bar{x}_{12},\bar{s}_d,\bar{x}_{1d},\bar{x}_{2d})^T
= {\bf P}\cdot {\bf W}^T$, where ${\bf P}$ is given by Eq.~(\ref{e:P}) and ${\bf
  W}=(W_1,W_2,W_3,W_4,W_5,W_6,W_7,W_8,W_9,W_{10})$,
 then the
transformation matrix leads to the following definition of new variables,
$W_i$, $i=1\ldots 10$:
\begin{equation}
\begin{split}
&W_1 = \frac{\bar{x}_{21}-\bar{x}_1}{1+\phi}, \quad W_2 =
\frac{\bar{x}_{12}-\bar{x}_2}{1+\phi}, \quad W_3 =
\frac{\bar{x}_{1d}+\bar{x}_{2d}-\bar{x}_1-\bar{x}_2}{\phi}, \quad W_4 =
\bar{x}_{2d}-\bar{x}_2, \\
&W_5 = \bar{s}, \quad W_6 =
\frac{\bar{x}_{1}+\phi\bar{x}_{21}}{1+\phi}, \quad W_7 =
\frac{\bar{x}_{2}+\phi\bar{x}_{12}}{1+\phi}, \quad W_8=\frac{\phi\bar{r}_{1}+\phi\bar{x}_{1}+\bar{r}_1-\phi\bar{x}_{21}}{1+\phi},\\
&W_9=\frac{\phi\bar{r}_{2}+\phi\bar{x}_{2}+\bar{r}_2-\phi\bar{x}_{12}}{1+\phi},\quad
W_{10} = \frac{\phi\bar{s}_{d}-\bar{x}_{1d}-\bar{x}_{2d}+\bar{x}_{1}+\bar{x}_2}{\phi}.\label{e:W1}
\end{split}
\end{equation}

\clearpage

\section{Transformed Evolution Equations}\label{sec:trans_evol_eq}

The application of the transformation matrix ${\bf P}$ given by Eq.~(\ref{e:P}) to
Eqs.~(\ref{e:dsbardtau})-(\ref{e:dx2dbardtau}) leads to the following set of
transformed evolution equations:
%\begin{footnotesize}
\begin{subequations}
\begin{flalign}
&\frac{dW_1}{d\tau} =\beta_0\left (W_6+\frac{\mu^2}{2\sigma_0}\right )\left
  (W_9+\phi W_2-W_5\right ) -\sigma_0 W_1 ,\label{e:dW1}\\
&\frac{dW_2}{d\tau} =\beta_0\left (W_7+\frac{\mu^2}{2\sigma_0}\right )\left
  (W_8+\phi W_1-W_5\right ) -\sigma_0 W_2 ,\label{e:dW2}\\
&\frac{dW_3}{d\tau} =W_3\left (\beta_0\left (W_3+W_{10}\right
  )-\frac{\sigma_0\phi}{1+\phi}\right )\nonumber \\
&\hspace{1cm}+\frac{\beta_0\left (1+\phi\right
  )}{\phi}\left (W_6+W_7+\frac{\mu^2}{\sigma_0}\right )\left
  (W_3+W_{10}-W_5\right ) ,\label{e:dW3}\\
&\frac{dW_4}{d\tau} =\left (\beta_0\left (W_3+W_{10}\right
  )-\frac{\sigma_0\phi}{1+\phi}\right )W_4 \nonumber \\
&\hspace{1cm}+\beta_0\left (W_3+W_{10}-W_5\right
)\left (\left (1+\phi\right )W_7+\frac{\mu^2\left (1+\phi\right )}{2\sigma_0}\right ),\label{e:dW4}\\
&\frac{dW_5}{d\tau} =\mu^2-\beta_0\left (W_5+\frac{\sigma_0}{\beta_0
    (1+\phi)}\right )\left ((1+\phi)\left (W_6+W_7+\frac{\mu^2}{\sigma_0}\right
    ) \right ) ,\label{e:dW5}\\
&\frac{dW_6}{d\tau} =\beta_0\left (W_6+\frac{\mu^2}{2\sigma_0}\right )\left
  (\phi^2 W_2+\phi W_9+W_5\right ) ,\label{e:dW6}\\
&\frac{dW_7}{d\tau} = \beta_0\left (W_7+\frac{\mu^2}{2\sigma_0}\right )\left
  (\phi^2 W_1+\phi W_8+W_5\right ),\label{e:dW7}\\
%\end{flalign}
%\begin{flalign}
&\frac{dW_8}{d\tau} =\left (W_6+\frac{\mu^2}{2\sigma_0}\right )\left
  (\sigma_0+\beta_0\left (\phi W_5 -\phi W_9 -\phi^2 W_2 \right ) \right )
\nonumber\\
&\hspace{1cm}-\left (1+\phi\right )\beta_0\left (W_7 + \frac{\mu^2}{2\sigma_0} \right )\left
  (\phi W_1 + W_8 + \frac{\sigma_0}{\beta_0\left (1+\phi\right )} \right ) ,\label{e:dW8}\\
&\frac{dW_9}{d\tau} =\left (W_7+\frac{\mu^2}{2\sigma_0}\right )\left
  (\sigma_0+\beta_0\left (\phi W_5 -\phi W_8 -\phi^2 W_1 \right ) \right )
\nonumber\\
&\hspace{1cm}-\left (1+\phi\right )\beta_0\left (W_6 + \frac{\mu^2}{2\sigma_0} \right )\left
  (\phi W_2 + W_9 + \frac{\sigma_0}{\beta_0\left (1+\phi\right )} \right ) ,\label{e:dW9}\\
&\frac{dW_{10}}{d\tau} =\mu^2-\left (1+\phi\right )\beta_0 W_3\left
  (W_3+W_{10}\right ) \nonumber \\
&\hspace{1cm}+ \frac{\left (1+\phi\right )\beta_0}{\phi}\left
  (W_6+W_7+\frac{\mu^2}{\sigma_0}\right )\left (W_5-\left (1+\phi\right )\left
    (W_3+W_{10}\right ) -\frac{\phi\sigma_0}{\beta_0\left (1+\phi\right )}\right ),\label{e:dW10}\\
&\frac{d\mu}{d\tau} =0. \label{e:dmu}
\end{flalign}
\end{subequations}

\clearpage

\section{Center Manifold Condition}\label{sec:CM-condition}

Substitution of the center manifold functions $W_i=h_i$ given by Eq.~(\ref{e:h1})
into the transformed evolution equations given in Appendix~\ref{sec:trans_evol_eq} 
leads to the following center manifold condition:
\begin{subequations}
\begin{flalign}
&\frac{\partial h_1}{\partial W_5}\frac{dW_5}{d\tau} + \frac{\partial
  h_1}{\partial W_6}\frac{dW_6}{d\tau} + \frac{\partial h_1}{\partial
  W_7}\frac{dW_7}{d\tau} + \frac{\partial h_1}{\partial W_8}\frac{dW_8}{d\tau}
+ \frac{\partial h_1}{\partial W_9}\frac{dW_9}{d\tau} + \frac{\partial
  h_1}{\partial W_{10}}\frac{dW_{10}}{d\tau} &\nonumber \\
&\hspace{1cm}=\beta_0\left (W_6+\frac{\mu^2}{2\sigma_0}\right )\left
  (W_9+\phi h_2-W_5\right ) -\sigma_0 h_1,\label{e:cmc-1}\\
&\frac{\partial h_2}{\partial W_5}\frac{dW_5}{d\tau} + \frac{\partial
  h_2}{\partial W_6}\frac{dW_6}{d\tau} + \frac{\partial h_2}{\partial
  W_7}\frac{dW_7}{d\tau} + \frac{\partial h_2}{\partial W_8}\frac{dW_8}{d\tau}
+ \frac{\partial h_2}{\partial W_9}\frac{dW_9}{d\tau} + \frac{\partial
  h_2}{\partial W_{10}}\frac{dW_{10}}{d\tau} \nonumber \\
&\hspace{1cm}=\beta_0\left (W_7+\frac{\mu^2}{2\sigma_0}\right )\left
  (W_8+\phi h_1-W_5\right ) -\sigma_0 h_2,\label{e:cmc-2}\\
&\frac{\partial h_3}{\partial W_5}\frac{dW_5}{d\tau} + \frac{\partial
  h_3}{\partial W_6}\frac{dW_6}{d\tau} + \frac{\partial h_3}{\partial
  W_7}\frac{dW_7}{d\tau} + \frac{\partial h_3}{\partial W_8}\frac{dW_8}{d\tau}
+ \frac{\partial h_3}{\partial W_9}\frac{dW_9}{d\tau} + \frac{\partial
  h_3}{\partial W_{10}}\frac{dW_{10}}{d\tau} &\nonumber \\
&\hspace{1cm}=h_3\left (\beta_0\left (h_3+W_{10}\right
  )-\frac{\sigma_0\phi}{1+\phi}\right )\nonumber \\
&\hspace{1cm}+\frac{\beta_0\left (1+\phi\right
  )}{\phi}\left (W_6+W_7+\frac{\mu^2}{\sigma_0}\right )\left
  (h_3+W_{10}-W_5\right ),\label{e:cmc-3}\\
&\frac{\partial h_4}{\partial W_5}\frac{dW_5}{d\tau} + \frac{\partial
  h_4}{\partial W_6}\frac{dW_6}{d\tau} + \frac{\partial h_4}{\partial
  W_7}\frac{dW_7}{d\tau} + \frac{\partial h_4}{\partial W_8}\frac{dW_8}{d\tau}
+ \frac{\partial h_4}{\partial W_9}\frac{dW_9}{d\tau} + \frac{\partial
  h_4}{\partial W_{10}}\frac{dW_{10}}{d\tau} &\nonumber \\
&\hspace{1cm}=\left (\beta_0\left (h_3+W_{10}\right
  )-\frac{\sigma_0\phi}{1+\phi}\right )h_4 \nonumber \\
&\hspace{1cm}+\beta_0\left (h_3+W_{10}-W_5\right
)\left (\left (1+\phi\right )W_7+\frac{\mu^2\left (1+\phi\right )}{2\sigma_0}\right ).\label{e:cmc-4}
\end{flalign}
\end{subequations}

\end{document}